\documentclass[reprint,aps,prl,twocolumn,groupedaddress,amsmath,amssymb]{revtex4-2}  % for review and submission
\usepackage{graphicx}  % needed for figures
\usepackage{dcolumn}   % needed for some tables
\usepackage{bm}        % for math
\usepackage{amssymb}   % for math
\usepackage{amsmath}   % for math
\usepackage{braket}    % Allows Dirac Notation
\usepackage{parskip}  % Turns OFF automatic indent
\usepackage{natbib}
\usepackage{hyperref}
\usepackage{subfig}
\usepackage{float}

\begin{document}

\title{Optimal Spin- and planar-quantum squeezing in superpositions of spin coherent states}
\author{Richard J. Birrittella$^{1}$, Jason Ziskind$^{2}$, Edwin E. Hach, III$^{2}$, Paul M. Alsing$^{1}$ and Christopher C. Gerry$^{3}$ \\
\textit{\textit{$^{1}$Air Force Research Laboratory, Information Directorate, Rome, NY, USA, 13441}\\
	$^{2}$Rochester Institute of Technology, School of Physics and Astronomy,\\
	85 Lomb Memorial Drive, Rochester, New York, 146232, USA}
	\\
\textit{$^{3}$Department of Physics and Astronomy, Lehman College,\\
The City University of New York, Bronx, New York, 10468-1589,USA} 
}

\date{\today}

\begin{abstract}
We investigate the presence of spin- and planar- squeezing in generalized superpositions of atomic (or spin) coherent states (ACS). Spin-squeezing has been shown to be a useful tool in determining the presence of entanglement in multipartite systems, such as collections of two-level atoms, as well as being an indication of reduced projection noise and sub-shot-noise limited phase uncertainty in Ramsey spectroscopy, suitable for measuring phases $\phi\sim 0$. On the other hand, planar-squeezed states display reduced projection noise in two directions simultaneously and have been shown to lead to enhanced metrological precision in measuring phases without the need for explicit prior knowledge of the phase value.  In this paper, we show that the generalized superposition state can be parametrized to display both spin-squeezing along all orthogonal axes and planar-squeezing along all orthogonal planes for all values of $J>1/2$. We close with an application of the maximally spin- and planar-squeezed states to quantum metrology.   
\end{abstract}

\maketitle

\section{\label{sec:Intro} I. Introduction}

\noindent Spin-squeezing is a non-classical effect that can be made to occur in ensembles of spin-$1/2$ particles, or in ensembles of two-level atoms, and corresponds to a  reduction of quantum noise in a chosen spin direction.  Entanglement between the spins is responsible for the spin-squeezing effect, serving as a means of determining entanglement-enhanced sensitivities in quantum sensing \cite{ref:Toth, ref:Pezze}, though not all entangled states display spin-squeezing. The practical use of spin-squeezing as an entanglement witness in large ensembles of atoms is of particular interest as the detection of multipartite entanglement in large systems is still at the forefront of quantum technologies development \cite{ref:Guhne2}. In the context of atomic systems, states that are spin-squeezed may be used for the quantum enhancement of measurements of transition frequencies between the atomic states \cite{ref:Wineland1, ref:Kitagawa}. On the other hand, states that are highly entangled but which contain no spin-squeezing are also suitable for this purpose.  For example, consider the maximally entangled states (MES) of the form $\ket{\psi_{\text{MES}}}=\tfrac{1}{\sqrt{2}}\left(\ket{J,J}+\ket{J,-J}\right)$ where the states $\ket{J,\pm J}$ are the Dicke states \cite{ref:Dicke} and where $N=2J$ is total number of atoms in the ensemble.  With these states, the uncertainty in the atomic frequency measurement $\delta\omega_{0}$ scales according to $\delta\omega_{0}=1/NT$ where $T$ is the free evolution time in the Ramsey spectroscopy procedure. Such a state, which yields Heisenberg-limited (HL) frequency sensitivity: an improvement over the standard quantum limit (SQL) by a factor of $\sqrt{N}$, raises an important distinction in how spin-squeezing is defined. While the MES display no spin-squeezing according to definitions put forth by S{\o}renson \textit{et al.} \cite{ref:Sorenson2}, who were interested in constructing a separability criterion for multipartite systems, it \textit{does} exhibit spin-squeezing as it is defined in the context of noise reduction in Ramsey spectroscopy by Wineland \textit{et al.} \cite{ref:Wineland1}. For the case of trapped ions, however, the MES are hard to make \cite{ref:Bollinger}, especially for larger ensembles of atoms in a gas. Despite this, the MES serve as an idealized example of a state capable of reaching the HL and can be considered the atomic analogue to the optical N00N state \cite{ref:Kok,ref:Dowling} of the form $\ket{\text{N00N}}\propto\left(\ket{N,0}+e^{i\theta}\ket{0,N}\right)$ in which the uncertainty in average photon number is equal to the total number of photons $\Delta N=N$ and the phase uncertainty is defined in regards to the heuristic uncertainty relation $\Delta\phi\Delta n \sim 1$ leading to the HL: $\Delta\phi \sim 1/N$.  On the other hand, recent work has been done in deriving a protocol for reaching the HL using a collective-state-detection scheme and utilizing a critically tuned one-axis twist Hamiltonian, leading to the generation of Sch{\"o}dinger-cat states \cite{ref:Sarkar}. The authors of \cite{ref:Sarkar} go on to show the narrowing of the interference fringes was critically dependent on the parity of the atomic ensemble, with a narrowing up to a factor of $N$. For large atomic ensembles, such as what one can generate via magneto-optical traps for neutral atoms, which can yield densities as high as $10^{12}$ atoms \cite{ref:Jagatap}, it is not possible to determine the atomic parity with certainty; however, when averaged over many runs in which the parity probabilities were equalized, they found a phase uncertainty nearing the HL up to a factor of $\sqrt{2}$. Work done by the same authors also demonstrated a means of reaching HL sensitivities in an atom interferometer with a Schr{\"o}dinger-cat state using increased quantum noise \cite{ref:Fang}, effectively suppressing by a factor of $\sqrt{N}$ the effect of excess noise. \\

\noindent Conversely, spin-squeezed states have been produced in various contexts such as cold atoms \cite{ref:Orzel, ref:Hosten}, trapped ions \cite{ref:Meyer}, magnetic systems \cite{ref:Auccaise} and photons \cite{ref:Mitchell} as well as discussed in the context of multi-atom atomic clocks \cite{ref:Schulte}\cite{ref:Fang} and room-temperature light-atom interactions in an optical cavity \cite{ref:Serafin} with a Helium-3 gas. Furthermore, schemes for generating entangled atomic ensembles in cavities have been proposed using a one- and two- axis twisting Hamiltonian \cite{ref:Kitagawa,ref:Molmer,ref:Borregaard}.  More recently, spin-squeezing has been experimentally demonstrated in an ensemble of $10^{11}$ atoms wherein the quantum state is generated via prediction and retrodiction quantum non-demolition measurements \cite{ref:Bao}.They report a squeezing of 4.5 decibels relative to a spin-coherent state and go on to demonstrate the practical use of their protocol with an application in atomic magnetometry. \\

\noindent One can also consider a form of squeezing in which one has a reduction in projection noise simultaneously in two orthogonal spin directions below the standard quantum limit $|\braket{\hat{J}_{\parallel}}|/2$, where $|\braket{\hat{J}_{\parallel}}|$ is the in-plane polarization, while increasing the noise in the third direction, known as planar quantum squeezing \cite{ref:He}.  Planar quantum squeezed (PQS) states have been demonstrated to yield enhanced phase sensitivity below the SQL for all phase angles, eliminating the need for \textit{a priori} knowledge of the phase.  Such states have proven useful for interferometric measurements involving tracking a moving phase and simultaneous phase-amplitude estimation below the standard quantum limit \cite{ref:Colangelo,ref:Vitagliano}. Planar squeezed states have been experimentally demonstrated through quantum non-demolition measures in cold atoms for spin-$1$ ensembles \cite{ref:Puentes}.\\

\noindent Some years ago, the non-classical properties of superpositions of atomic coherent states were explored \cite{ref:Gerry}\cite{ref:Puri} in detail.  The atomic coherent states are given by \cite{ref:Arrechi}

\begin{equation}
	\ket{\zeta,J} = \left(1+|\zeta|^{2}\right)^{-J}\sum_{M=-J}^{J}\binom{2J}{J+M}^{1/2}\zeta^{J+M}\ket{J,M},
	\label{eqn:Intro_1}
\end{equation}

\noindent where $\zeta=e^{i\phi}\tan\tfrac{\theta}{2}$ and where $0\leq\theta\leq\pi,\;0\leq\phi\leq 2\pi$ parametrize the Bloch sphere.  The superposition states considered by Gerry and Grobe \cite{ref:Gerry} are given by

\begin{equation}
	\ket{\Psi_{\pm}} = \mathcal{N}_{\pm}\left(\ket{\zeta,J} \pm e^{-i\pi J}\ket{-\zeta,J}\right),
	\label{eqn:Intro_2}
\end{equation}

\noindent where the normalization factor $\mathcal{N}_{\pm}$ is 

\begin{equation}
	\mathcal{N}_{\pm} = \tfrac{1}{\sqrt{2}}\left[1+\cos\left(\delta_{\pm}-\pi J\right)\left(\frac{1-|\zeta|^{2}}{1+|\zeta|^{2}}\right)^{2J}\right]^{-1/2},
	\label{eqn:Intro_3}
\end{equation}

\noindent with $\delta_{+}\equiv 0\;\text{and}\;\delta_{-}\equiv \pi$.  States for which $J\in\{2k : k\in \mathbb{Z}^{0+}\}$ are analogous to the even coherent states of the usual harmonic oscillator based states while $J\in\{2k+1 : k\in \mathbb{Z}^{0+}\}$ are the analogs of the odd coherent states and finally the case where $J\in \{\left(2k+1\right)/2 : k\in \mathbb{Z}^{0+}\}$ are the analogs of the Yurke-Stoler (Y-S) states \cite{ref:Yurke2, ref:Sanders}.  The states of Eq.~\ref{eqn:Intro_2} are entangled and shown to exhibit spin-squeezing as defined by Wineland \textit{et al.}\cite{ref:Wineland1}.  \\

\noindent In this paper we consider the more general class of superposition states, having the form 

\begin{equation}
	\ket{\Psi} = \mathcal{N}_{J}\left(\ket{\zeta_{1},J} + e^{i\Phi_{r}}\ket{\zeta_{2},J}\right),
	\label{eqn:Intro_4}
\end{equation}

\noindent where 

\begin{equation}
	\mathcal{N}_{J} = \frac{1}{\sqrt{2}}\Bigg(1 + \frac{\text{Re}\left[\left(1+\zeta_{1}^{*}\zeta_{2}\right)^{2J}e^{i\Phi_{r}}\right]}{\left(1+|\zeta_{1}|^{2}\right)^{J}\left(1+|\zeta_{2}|^{2}\right)^{J}}\Bigg)^{-1/2},
	\label{eqn:Intro_5}
\end{equation}

\noindent and where $\zeta_{i} = e^{i\phi_{i}}\tan\tfrac{\theta_{i}}{2}$. The spin state given in Eq.~\ref{eqn:Intro_4} is a general linear superposition of atomic coherent states localized around two (potentially widely separated) positions on the Bloch sphere. This choice is motivated, in part, by the work of Schlaufler \textit{et al.} \cite{ref:Schaufler} who studied, in the context of a single-mode quantized field, what they called a phase-cat state; that is, a superposition of two Glauber coherent states \cite{ref:Glauber1} of different phases but with identical amplitudes.  This work is also motivated by the works of Prakash and Kumar \cite{ref:Prakash}\cite{ref:Prakash1} who studied similar superpositions of Glauber states but where the amplitudes and phases were allowed to vary. \\

\noindent In terms of the Dicke states, the state $\ket{\Psi}$ of Eq.~\ref{eqn:Intro_4} can be expanded accordingly as

\begin{equation}
	\ket{\Psi} = \sum_{M=-J}^{J} C_{M}^{\left(J\right)}\left(\theta_{1},\theta_{2},\phi,\Phi_{r}\right)\ket{J,M},
	\label{eqn:Prop_1}
\end{equation}

\noindent where the expansion coefficients are given by 

\begin{align}
	C_{M}^{\left(J\right)}&\left(\theta_{1},\theta_{2},\phi,\Phi_{r}\right) = \mathcal{N}_{J}\binom{2J}{J+M}^{1/2}\times \nonumber \\
	&\;\;\;\;\;\;\;\;\;\;\;\;\;\times \left(S_{1}^{J}T_{1}^{M} + e^{i\left[\Phi_{r}+ \left(J+M\right)\phi\right]}S_{2}^{J}T_{2}^{M}\right), 
	\label{eqn:Prop_2}
\end{align}

\noindent and where the designations $S_{i}\equiv\sin\theta_{i}$ and $T_{i}\equiv \tan\tfrac{\theta_{i}}{2}$ have been made.  The normalization factor $\mathcal{N}_{J}$ is given by

\begin{align}
	\mathcal{N}_{J} &= \left[\sum_{M=-J}^{J}\binom{2J}{J+M}\left(S_{1}^{2J}T_{1}^{2M} + S_{2}^{2J}T_{2}^{2M} + \right.\right. \nonumber \\
	&  \left.\left. + 2\left(S_{1}S_{2}\right)^{J}\left(T_{1}T_{2}\right)^{M}\cos\left[\Phi_{r} + \left(J+M\right)\phi\right]  \right)\right]^{-1/2}.
	\label{eqn:Prop_3}
\end{align}

\noindent In what follows, and without loss of generality, we set $\phi_{1} = 0\;\text{and}\;\phi_{2}=\phi$ so that $\zeta_{1} = \tan\tfrac{\theta_{1}}{2}$ and $\zeta_{2} = e^{i\phi}\tan\tfrac{\theta_{2}}{2}$.  Our goal is to treat the parameters $\phi,\;\theta_{1},\;\theta_{2}\;\text{and}\;\Phi_{r}$ as variables over which we optimize to show maximal spin-squeezing and planar-squeezing for a particular value of collective spin value $J$ .  As will be shown below, optimal spin-squeezing generally does not occur for the spin-states analogous to the even, odd or Y-S states of the quantized field, though phase-cat states \cite{ref:Buzek} often optimize spin-squeezing.  Further, we show that for the correct choice of state parameters, significant planar squeezing exists in this system. \\

\noindent This paper is organized as follows: In Sec. \hyperref[sec:SSnPQS]{II\ref*{sec:SSnPQS}} we provide a brief review on the different forms of spin-squeezing used in the literature, most notably the definitions put forth by Wineland \textit{et al.} \cite{ref:Wineland1} and S{\o}renson \textit{et al.} \cite{ref:Sorenson2} with respect to metrological precision and entanglement, respectively. We also introduce planar squeezing as defined by He \textit{et al.} \cite{ref:He} and the connection between planar squeezing and the depth of entanglement of the ensemble.  In all cases, we include plots of the optimized spin- and planar squeezing along with a table of parameters optimizing the squeezing.  In Sec. \hyperref[sec:Metrology]{III\ref*{sec:Metrology}} we apply the optimally spin- and planar- squeezed states to atom interferometry in which we show the optimally spin squeezed state, corresponding to an even cat state yields greater phase sensitivity for small phases while the optimally planar squeezed state, closely related to a phase-cat state, yields sub-SQL phase uncertainty for a slightly broader range of phase.  We close in Sec. \hyperref[sec:SpinSqueeze_Conclusion]{IV\ref*{sec:SpinSqueeze_Conclusion}} with a discussion summarizing our findings. \\

\begin{figure}
	\centering
	\subfloat[][]{\includegraphics[width=0.985\linewidth,keepaspectratio]{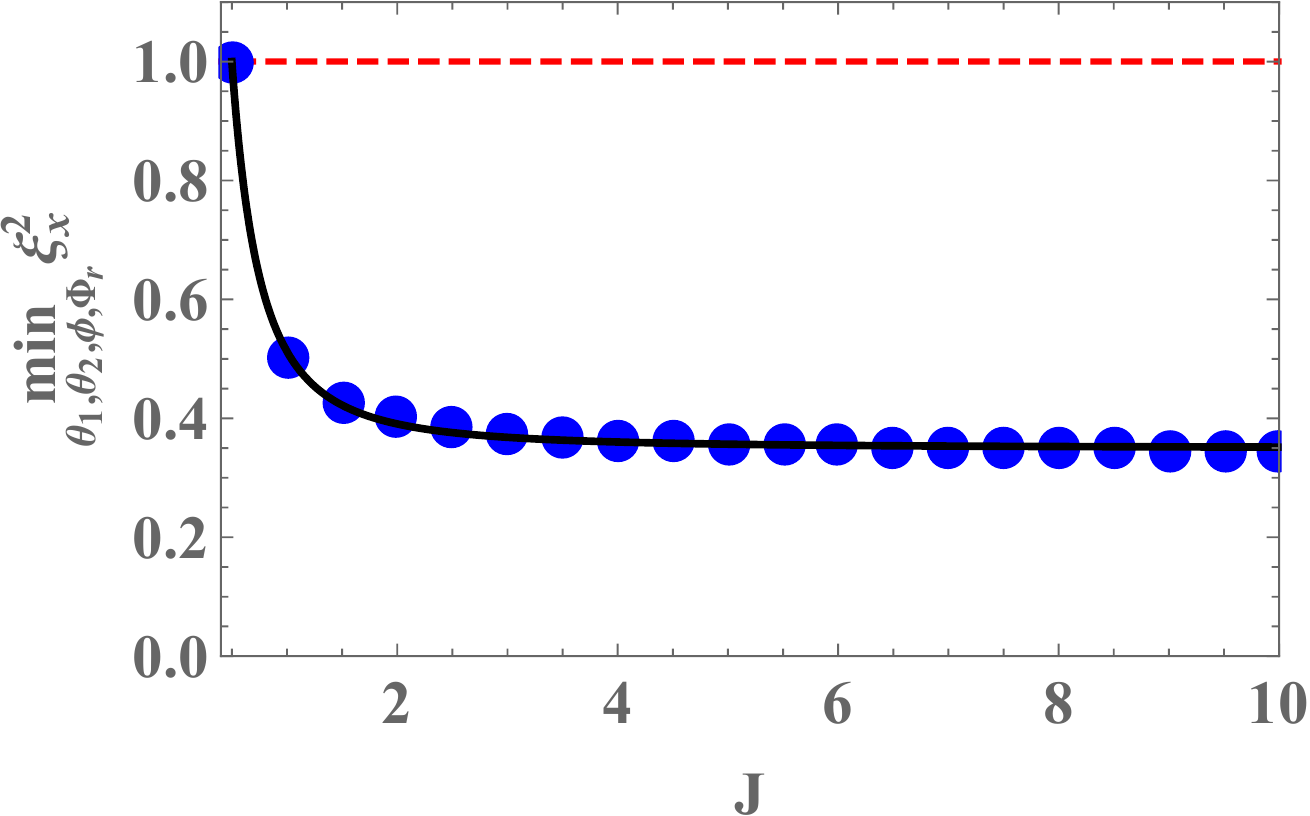}
		\label{fig:SSx}}
	\\
	\subfloat[][]{\includegraphics[width=0.985\linewidth,keepaspectratio]{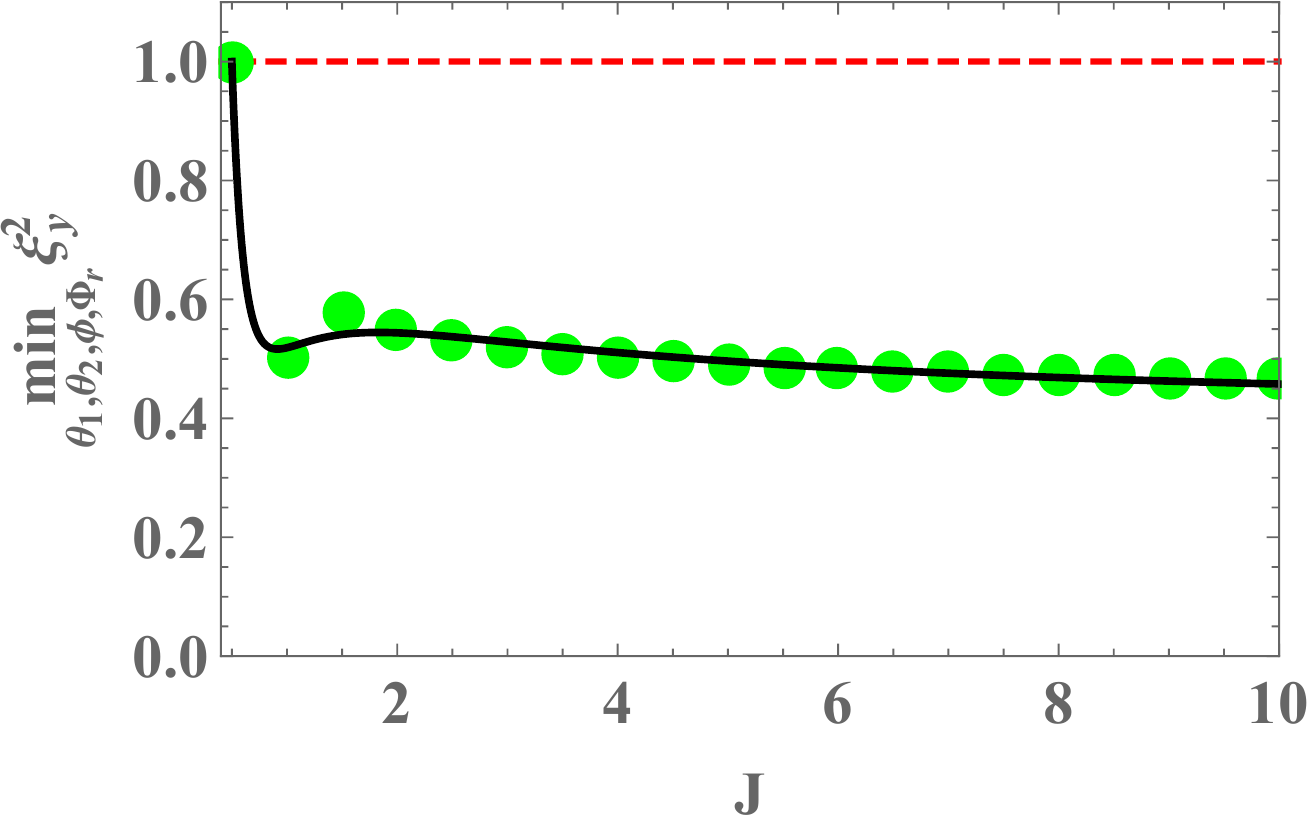}
		\label{fig:SSy}}
	\\
	\subfloat[][]{\includegraphics[width=0.985\linewidth,keepaspectratio]{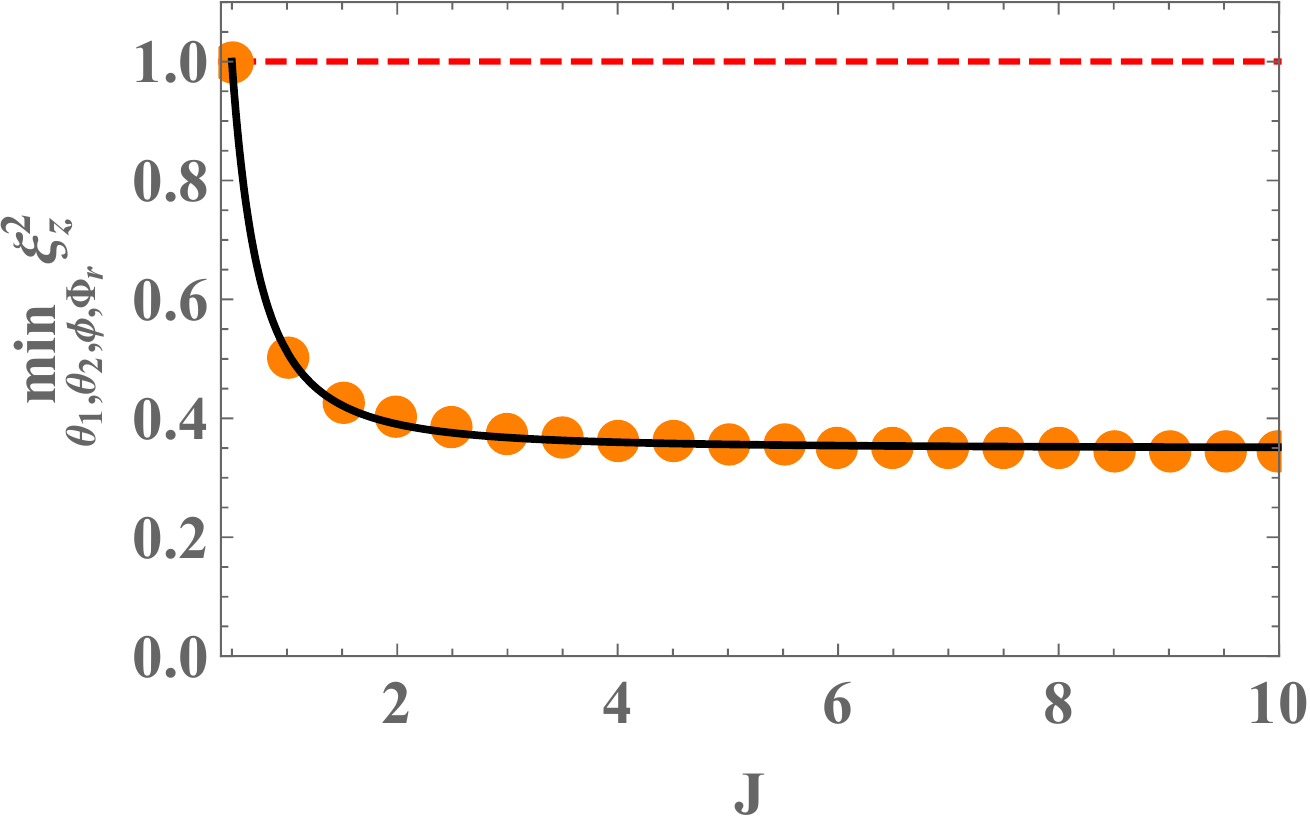}
		\label{fig:SSz}}
	\caption{Minimum spin-squeeze parameter $\xi_{i}^{2}$ for the \ref{fig:SSx} $x$-spin direction, \ref{fig:SSy} $y$-spin direction and \ref{fig:SSz} $z$-spin direction, respectively.  For each, a corresponding approximate curve-fit (red solid line) is included.}
	\label{fig:SS_plots} 
\end{figure}

\section{\label{sec:SSnPQS} II. Optimized Spin- and Planar- Squeezing}

\begin{table*}
	\caption{\label{tab:table_SS} Minimum values of $\xi_{k}^{2},\;k=x,y,z$ along with the corresponding state parameters for increasing values of total spin $J$.}
	\begin{ruledtabular}
		\begin{tabular}{cccccccc}
			$J$&$\xi_{x}^{2}$&$\xi_{y}^{2}$&$\xi_{z}^{2}$&$\theta_{1}$&$\theta_{2}$&$\phi$&$\Phi_{r}$\\
			\hline
			&&&&&&&\\
			1/2 & \textbf{1}&-&-& 1.4784 & 1.47775 & 6.20295 & 0.634272   \\
			& -&\textbf{1}&-& 0.0514062 & 0.0675389 & 1.37728 & 0.08810551  \\
			& -&-&\textbf{1}& 3.05204 & 3.1233 & 0.0913789 & 6.24518  \\
			1 &  \textbf{0.5}&-&-& 1.55444 & 1.57172 & 0.0163226 & 3.12513 \\
			& -&\textbf{0.5}&-	& 1.12713 & 2.01633 & 3.14159 & 6.28319   \\
			& -&-&\textbf{0.5}& 1.48475 & 1.48475 & 3.14272 & 3.14057  \\
			3/2 &  \textbf{0.428602}&-&-& 1.56079 & 1.57234 & 0.0099 & 3.12666  \\
			&-&\textbf{0.5802}&-&1.748 & 0.001 & 6.20574 & 0.01  \\
			& -&-&\textbf{0.428612}& 0.0145656 & 0.0145655 & 2.06606 & 3.14173 \\
			2 & \textbf{0.400095}&-&-& 1.55125 & 1.57999 & 0.017252 & 3.10659 \\
			& -&\textbf{0.550874}&-& 3.14159 & 1.63663 & 6.28144 & 0.01   \\
			& -&-&\textbf{0.400024}& 0.01 & 0.01 & 2.11057 & 3.15168  \\
			5 &  \textbf{0.357216}&-&-& 1.55977 & 1.57551 & 0.00997 & 3.09137 \\
			& -&\textbf{0.489156}&-& 3.14159 & 2.16809 & 6.28228 & 0.01 \\
			& -&-&\textbf{0.357208}& 0.01 & 0.01 & 1.93091 & 3.14183  \\
			10 &  \textbf{0.345018}&-&-& 1.5583 & 1.57549 & 0.01158 & 3.02478 \\
			& -&\textbf{0.466658}&-& 3.14159 & 2.44198 & 6.2827 & 0.01  \\
			& -&-&\textbf{0.344977}& 0.01052 & 0.01052 & 1.86283 & 3.14212  \\
			\label{tab:SStable}
		\end{tabular}
	\end{ruledtabular}
\end{table*}

\subsection{\label{sec:SS} a. Spin-Squeezing}

\noindent Spin-squeezing does not have a unique definition.  It has been defined and used by many different authors within several contexts. Some of the early definitions of spin-squeezing were derived directly from the Robertson uncertainty relation governing spin-operators $\Delta\hat{J}_{i}\Delta\hat{J}_{j}\geq\tfrac{1}{4}|\braket{\hat{J}_{k}}|$, where $\big[\hat{J}_{i},\hat{J}_{j}\big]=i\epsilon_{ijk}\hat{J}_{k}$ and where $i,j,k$ represent orthogonal directions.  If the mean spin direction (MSD) lies within the $i-j$ plane, then the consequence of the uncertainty relation is that $\Delta\hat{J}_{i\left(j\right)}>0$.  The variances can be made arbitrarily small with the uncertainty being absorbed in the orthogonal direction.  Definitions of spin-squeezing with respect to this uncertainty relation are undesirable because they are not defined with respect to a classical limit to overcome.  Consequently, states which do not display any quantum properties, such as the spin- (or atomic-) coherent states, \textit{can} display a form of 'spin-squeezing': specifically, spin-squeezing defined with respect to the angular momentum uncertainty relations.  We refer the reader to a paper by Ma \textit{et al.} \cite{ref:Ma} which provides a comprehensive history of the different forms of spin-squeezing.  For our purposes, however, we consider two closely related definitions of spin-squeezing put forth by S{\o}renson \textit{et al.} \cite{ref:Sorenson2} and by Wineland \textit{et al.} \cite{ref:Wineland1}. The former constructed a parameter for detecting spin-squeezing along a particular axis, defined arbitrarily along the $\vec{n}_{1}$-axis by $\xi^{2} = N\braket{\big(\Delta\hat{J}_{\vec{n}_{1}}\big)^{2}}/\big(\braket{\hat{J}_{\vec{n}_{2}}}^{2} + \braket{\hat{J}_{\vec{n}_{3}}}^{2}\big)$, where $\vec{n}_{i}$ represent general orthogonal directions and $\hat{J}_{\vec{n}_{i}}=\vec{J}\cdot\vec{n}_{i}$. For the $k$-axis, $k=x,y,z$, this is written compactly as

\begin{equation}
	\xi_{k}^{2} = \frac{N\braket{\big(\Delta\hat{J}_{k}\big)^{2}}}{\braket{\hat{J}_{k \perp}}^{2}},
	\label{eqn:Intro_6}
\end{equation}

\noindent where $N=2J$ and where $\braket{\hat{J}_{\perp k}}$ represents the mean spin in the plane perpendicular to the $k$-axis. The motivation for this came about by finding a condition of separability for spin-$1/2$ ensembles.  That is, assuming a separable density matrix of the form $\rho_{N}=\sum_{k}P_{k}\bigotimes_{i=1}^{N}\rho_{i}^{\left(k\right)}$ necessitates the condition $\xi_{k}^{2} \geq 1$. Therefore, spin-squeezing implies some degree of entanglement; however, states can be entangled \textit{without} being spin-squeezed. A ``one-way'' means of detecting entanglement being indicative of an entanglement witness. This form of spin-squeezing is closely related to the definition of spin-squeezing defined by Wineland \textit{et al.} \cite{ref:Wineland1} with regards to Ramsey spectroscopy

\begin{equation}
	\xi^{2}_{W} = \frac{\left(\Delta\varphi\right)^{2}}{\left(\Delta\varphi_{\text{SQL}}\right)^{2}}= \frac{N\braket{\big(\Delta\hat{J}_{\vec{n}\perp}\big)^{2}}}{|\braket{\vec{J}}|^{2}},
	\label{eqn:Intro_6a}
\end{equation}

\noindent where $\vec{n}_{\perp}$ is the unit vector perpendicular to the MSD,  $\hat{J}_{\vec{n}_{\perp}}=\vec{J}\cdot\vec{n}_{\perp}$, and where $\Delta\varphi_{\text{SQL}}$ represents the SQL of phase sensitivity obtained for the atomic coherent states of Eq.~\ref{eqn:Intro_1}.  The forms of spin-squeezing given in Eqs.~\ref{eqn:Intro_6} and \ref{eqn:Intro_6a} are closely related given that if $\vec{n}_{1}$ is chosen to minimize the variance $\Delta\hat{J}_{\vec{n}_{1}}$ while the MSD lies along the $\vec{n}_{2}$-axis, these expressions are equivalent.  In this respect, Eq.~\ref{eqn:Intro_6} can be considered a generalization of Eq.~\ref{eqn:Intro_6a} \cite{ref:Ma}.  Spin-squeezing along a particular direction exists whenever one has $\xi_{W}^{2} < 1$.  As these definitions of spin-squeezing are related to a classical limit (i.e. the SQL of phase uncertainty for the case of the Wineland \textit{et al.} \cite{ref:Wineland1} spin-squeeze parameter and a condition of separability for the case of the S{\o}renson \textit{et al.} \cite{ref:Sorenson2} spin-squeeze parameter), the spin- (or atomic-) coherent state will not be spin-squeezed. For the S{\o}renson \textit{et al.} \cite{ref:Sorenson2} definition of spin-squeezing, the spin-squeeze parameter is $\xi_{k}^{2}\equiv 1\; \forall\; k$ when calculated for the ACS while for the Wineland \textit{et al.} \cite{ref:Wineland1} definition of spin-squeezing, the spin-squeeze parameter yields $\xi_{W}^{2}\geq 1$ in the same case, dependent on the choice of phase being measured.  That is, when defining spin-squeezing with respect to a classical bound, the ACS does not exhibit spin-squeezing: the state itself is not entangled, by definition, and cannot yield sub-SQL phase sensitivity.  It should be noted that it is not possible for a state to exhibit spin-squeezing in three orthogonal directions simultaneously as a result of the uncertainty relation governing angular momentum, however it is possible for a state to be spin-squeezed in two directions simultaneously.  \\

\noindent In Fig.~\ref{fig:SS_plots} we plot the minimum spin-squeeze parameter in the $x,y,z$-directions as a function of collective spin $J$ and provide a table of parameters minimizing these spin-squeeze parameters in Table~\ref{tab:SStable} for up to $N=2J=10$ atoms. The spin-squeeze parameters are optimized across state parameters $\theta_{1},\;\theta_{2},\;\Phi_{r},\;\text{and}\;\phi$ defined in terms of the generalized superposition state above.  For each data set, we include an approximate estimation of the curve. In the $x,z$-directions the state minimizing the spin-squeeze parameter closely correspond to a phase-cat state with some relative phase of the form $\ket{\psi_{\text{p-cat}}} \propto \ket{\tau,J}+e^{i\theta}\ket{\tau e^{i\phi},J}$. For both cases, the curve fit (red solid line in Figs.~\ref{fig:SSx} and \ref{fig:SSz}) is of the form $a+bx^{2}$ where $x\equiv1/J$, $a\sim 0.35$ and $b\sim 0.16$. The state minimizing $\xi_{y}^{2}$ does not correspond to a cat nor a phase-cat state (see Tab.~\ref{tab:SStable}). The curve fit in Fig.~\ref{fig:SSy} is of the form $\sum_{i=0}^{3}a_{i}x^{i}$ with $x=1/J$ and $a_{0,1,2,3}\sim 0.4,0.61,0.84,0.34$.     

\subsection{\label{PQS} b. Planar Quantum Squeezing}

\noindent As we have discussed above, spin-squeezing as defined by S{\o}renson \textit{et al.} \cite{ref:Sorenson2} and given by Eq.~\ref{eqn:Intro_6}, can serve as a suitable means of detecting multipartite entanglement in atomic ensembles. More specifically, the condition $\xi_{k}^{2}<1$ necessitates the corresponding $N$ two-level atomic ensemble ($j=1/2$) cannot be written as a separable state. S{\o}renson \textit{et al.} \cite{ref:Sorenson1} generalize their findings for spin-$j>1/2$ and introduce a ``depth'' of entanglement corresponding to the number of atoms comprising the largest separable subset of the system. \\

\noindent However, one can investigate the entanglement properties of an ensemble of atoms using a different metric for which one can quantify a depth of entanglement: planar quantum squeezing (PQS). PQS was first introduced by He \textit{et al.} \cite{ref:He} who discussed a lower bound on a sum of spin variances in a plane, i.e. $\Delta^{2}\vec{J}_{\parallel}=\Delta^{2}\hat{J}_{i}+\Delta^{2}\hat{J}_{i\perp}\geq C_{J}$ where $C_{J}$ are tabulated lower bounds with a fractional exponent scaling $C_{J}\sim J^{2/3}$. They went on to show that states for which this inequality is saturated display the same fractional exponential scaling $\Delta^{2}\vec{J}_{\parallel}\sim J^{2/3}$, while the spin perpendicular to the plane has scaling $\Delta^{2}\vec{J}_{\perp}\sim J^{4/3}$. From this, entanglement among $N=2J$ sites of a spin-$j$ ensemble can be determined from the criterion $\Delta^{2}\vec{J}_{\parallel}^{\text{coll.}}<NC_{J}$, where $\vec{J}_{\parallel}^{\text{coll.}}$ represents the collective $N$-atom spin operator.  As stated previously, states that are planar squeezed display reduced spin variances in two perpendicular directions in a plane; however, this leads to the uncertainty being absorbed by the third spin component as per the uncertainty relation governing spin angular momentum.   \\

\noindent Choosing for example the $x-y$ plane such that we have the uncertainty relation $\Delta\hat{J}_{x}\Delta\hat{J}_{y}\geq \tfrac{1}{2}|\braket{\hat{J}_{z}}|$, one can form a functional definition of planar quantum squeezing by defining the planar variance as $\Delta^{2}\vec{J}_{\parallel}=\Delta^{2}\hat{J}_{x} + \Delta^{2}\hat{J}_{y}$ having large in-plane mean spin $|\braket{\vec{J}_{\parallel}}|=\sqrt{\braket{\hat{J_{x}}}^{2}+\braket{\hat{J_{y}}}^{2}}$. The SQL governing the spin variances forming the plane are taken to be $\Delta^{2}\hat{J}_{x}=\Delta^{2}\hat{J}_{y}=\tfrac{1}{2}|\braket{\vec{J}_{\parallel}}|$, while the SQL for the planar variance is given by $\Delta^{2}\vec{J}_{\parallel}=|\braket{\vec{J}_{\parallel}}|$.  These limits form a pseudo-'classical bound' for which we characterize a planar squeezed state. A planar squeezing parameter \cite{ref:Colangelo} can be defined as

\begin{equation}
	\xi_{xy}^{2} = \frac{\Delta^{2}\vec{J}_{\parallel}}{|\braket{\vec{J}_{\parallel}}|} = \frac{\Delta^{2}\hat{J}_{x} + \Delta^{2}\hat{J}_{y}}{\sqrt{\braket{\hat{J_{x}}}^{2}+\braket{\hat{J_{y}}}^{2}}}\equiv \xi_{\parallel}^{2},
	\label{eqn:PQS1}
\end{equation}

\noindent in which a planar squeezed state satisfies $\xi_{\parallel}^{2}<1$.  With this operational definition of planar squeezing, it is possible to parameterize an atomic coherent state to display planar squeezing; a state that is not entangled. In this sense, planar squeezing is easier to achieve than entanglement \cite{ref:Puentes}.  Consequently, unlike the definitions of spin-squeezing discussed in the previous section (as per S{\o}renson \textit{et al.} \cite{ref:Sorenson2} and Wineland \textit{et al.} \cite{ref:Wineland1}), planar squeezing \textit{does not} necessitate entanglement.  A ``depth of entanglement" was defined by Vitagliano \textit{et al.} \cite{ref:Vitagliano}, based on the condition 

\begin{equation}
	\xi_{\parallel}^{2} \geq D_{J}
	\label{eqn:PQS2}
\end{equation}

\noindent where $D_{J}$ is the minimum value of the planar squeeze parameter over single particle states of spin-$J$, tabulated by previous authors \cite{ref:Vitagliano}.  It was shown that for spin-$j$ ensembles, the condition in Eq.~\ref{eqn:PQS2} implies the ensemble contains $k$-entangled particles \cite{ref:Guhne} \textit{at most}, where $J=kj$. Likewise the condition $\xi_{\parallel}^{2} < D_{j}$ implies a depth of entanglement of \textit{at least} $\left(k+1\right)$. For example, consider once again the atomic coherent state.  For this state, the smallest value that $\xi_{\parallel}^{2}$ can take is $0.5$, indicating the presence of planar squeezing.  However, the largest value tabulated by Vitagliano \textit{et al.} \cite{ref:Vitagliano} for $D_{J}$ is $0.45$, confirming that entanglement cannot be discerned through planar squeezing.   This is sensible as the atomic coherent state is a separable state by definition. This raises a general distinction between SS and PQS states: SS is a sufficient criteron to show the possibility of sub-SQL sensitivity in interferometric measurements as the presence of SS requires entanglement within the system. However, the same is not true of PQS: PQS does not necessitate entanglement within the ensemble. Entanglement remains a crucial element in achieving sensitivity below the shot-noise limit.\\

\begin{figure}
	\centering
	\subfloat[][]{\includegraphics[width=0.985\linewidth,keepaspectratio]{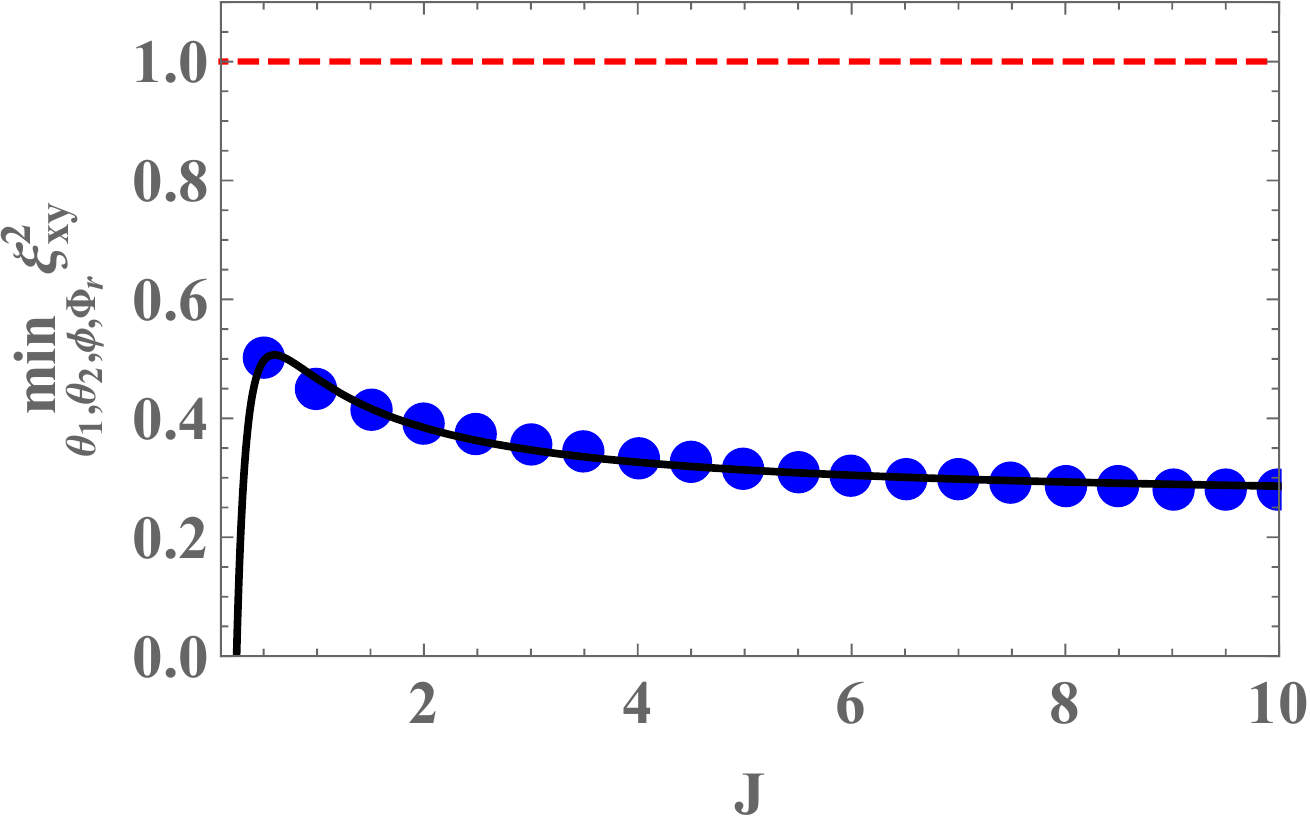}
		\label{fig:PQSxy}}
	\\
	\subfloat[][]{\includegraphics[width=0.985\linewidth,keepaspectratio]{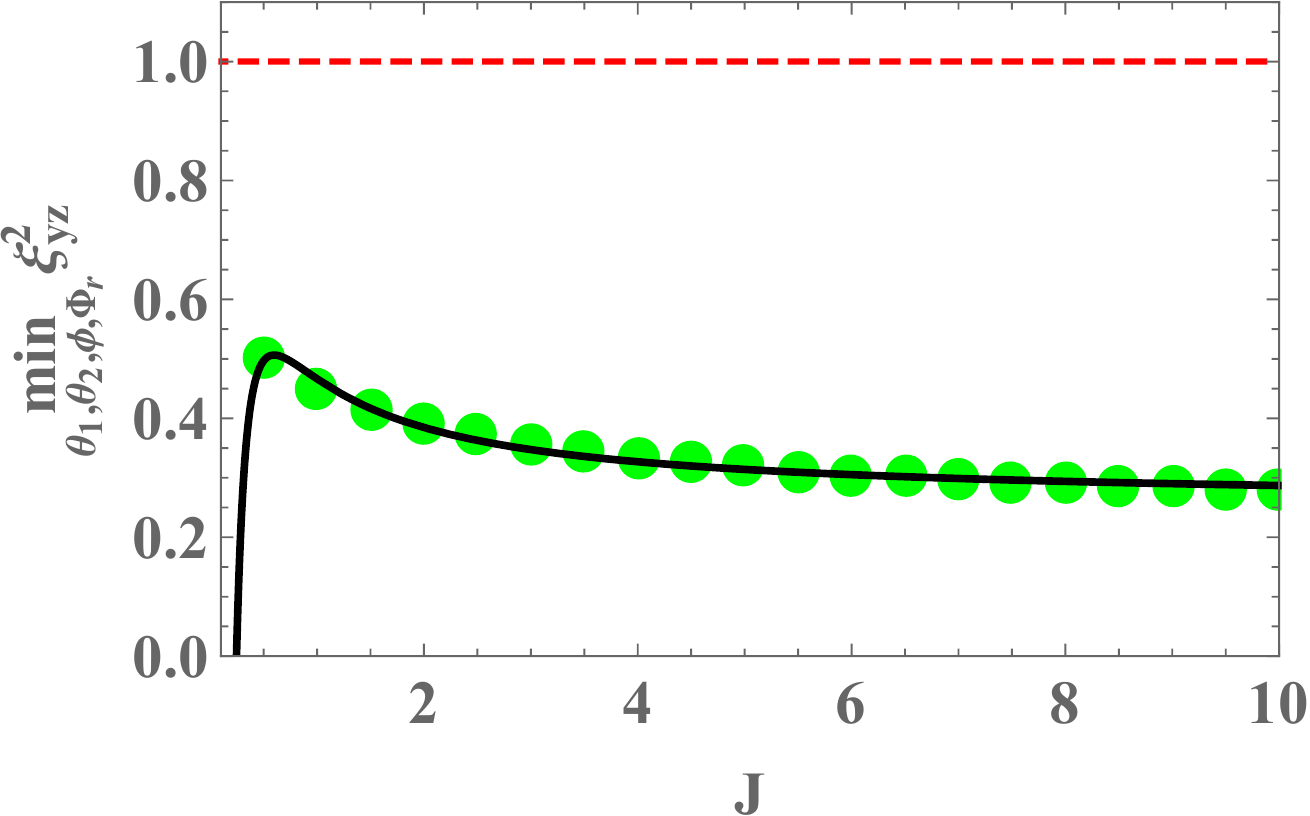}
		\label{fig:PQSyz}}
	\\
	\subfloat[][]{\includegraphics[width=0.985\linewidth,keepaspectratio]{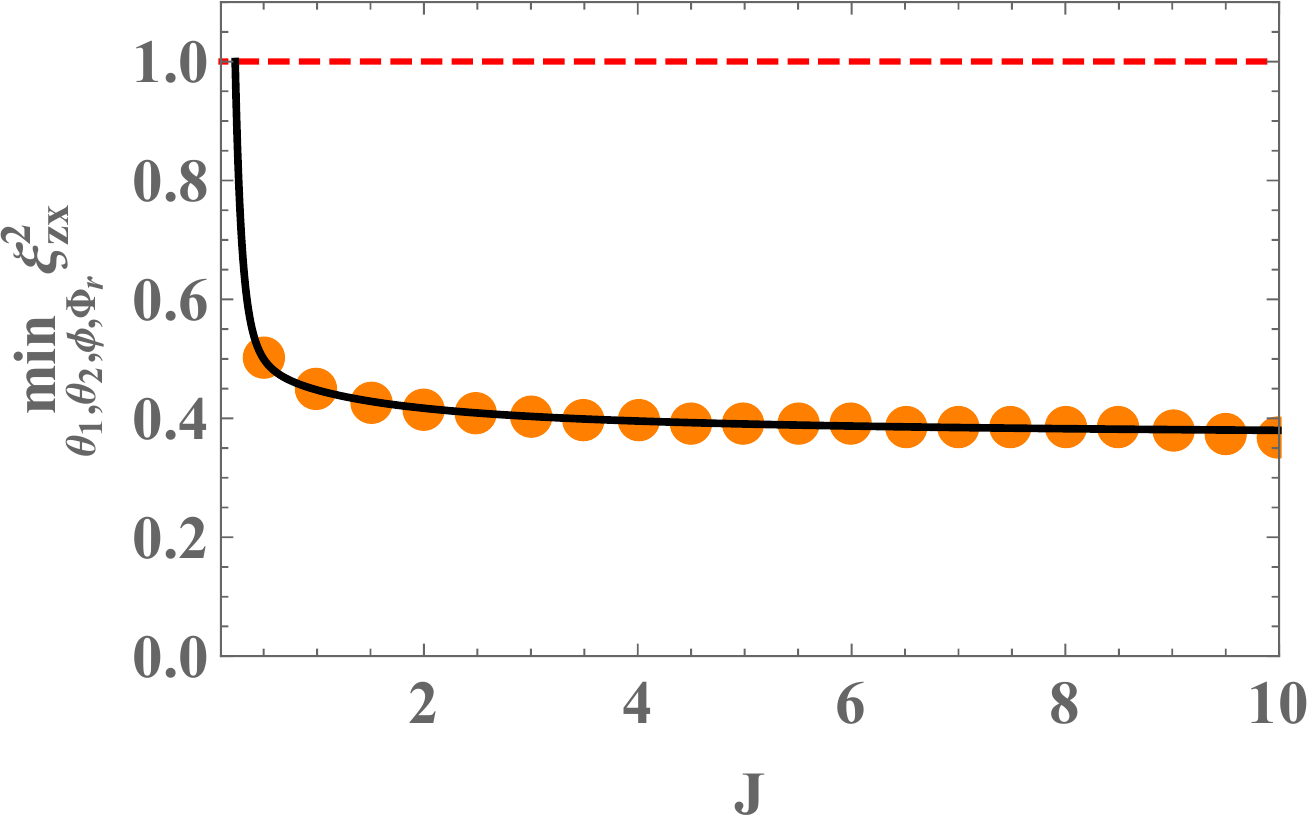}
		\label{fig:PQSzx}}
	\caption{Minimum planar squeeze parameter $\xi_{ij}^{2}$ for the \ref{fig:PQSxy} $x-y$ plane, \ref{fig:PQSyz} $y-z$ plane and \ref{fig:PQSxy} $z-x$ plane, respectively. For each, a corresponding approximate curve-fit (red solid line) is included.}
	\label{fig:PQS_plots} 
\end{figure}

\noindent Plots of the minimum planar squeeze parameter $\xi_{ij}^{2}$ where $ij=xy,yz,zx$ are provided in Fig.~\ref{fig:PQS_plots} along with corresponding state parameters in Tab.~\ref{tab:table_PQS} up to $N=2J=10$. The states corresponding to maximal planar squeezing in the $y-z$ and $z-x$ planes are approximately phase cat states, while the state corresponding to maximal planar squeezing in the $x-y$ plane are neither cat states nor phase cat states. Once again we include an approximate curve fit where for the $x-y$ and $y-z$ planes has the form $a+bx+cx^{2}$ with $a\sim 0.257,\;b\sim0.089,\;\text{and}\;c\sim0.3$ where $x\equiv1/J$.  For the $z-x$ plane, the curve fit in Fig.~\ref{fig:PQSzx} is of the form $\sum_{i=0}^{3}a_{i}x^{i}$ with $a_{0,1,2,3}\sim 0.37,\;0.12,\;-0.06,\;0.015$.

\begin{table*}
	\caption{\label{tab:table_PQS} Minimum values of $\xi_{ij}^{2},\;ij=xy,yz,zx$ along with the corresponding state parameters for increasing values of total spin $J$.}
	\begin{ruledtabular}
		\begin{tabular}{cccccccc}
			$J$&$\xi_{xy}^{2}$&$\xi_{yz}^{2}$&$\xi_{zx}^{2}$&$\theta_{1}$&$\theta_{2}$&$\phi$&$\Phi_{r}$\\
			\hline
			&&&&&&&\\
			1/2 & \textbf{0.5}&-&-& 3.07749 & 0.393397 & 4.62433 & 3.57279 \\
			& -&\textbf{0.5}&-& 0.175436 & 0.198689 & 3.75173 & -0.160262\\
			& -&-&\textbf{0.5}& 0.653902 & 0.767073 & 6.27056 & 0.156313 \\
			1 &  \textbf{0.44906}&-&-& 2.09702 & 1.04457 & 6.28319 & 1.2433 \\
			& -&\textbf{0.44906}&-	& 0.589433 & 0.589433 & 3.14159 & -0.719954  \\
			& -&-&\textbf{0.449065}& 0.671033 & 0.671092 & 6.27062 & 3.13338\\
			3/2 &  \textbf{0.414836}&-&-& 2.14272 & 0.998877 & 6.28318 & 0.0001 \\
			&-&\textbf{0.414836}&-& 0.571919 & 0.571919 & 3.14159 & 5.669e-6\\
			& -&-&\textbf{0.427156}& 0.631818 & 0.631852 & 0.01 & 3.14838 \\
			2 & \textbf{0.389929}&-&-& 2.10559 & 1.03601 & 6.28318 & 0.0001 \\
			& -&\textbf{0.389929}&-& 0.534789 & 0.534789 & 3.14159 & 4.319e-8 \\
			& -&-&\textbf{0.415149}& 0.682875 & 0.68289 & 0.01 & 3.12543 \\
			5 &  \textbf{0.317381}&-&-& 1.97969 & 1.1619 & 6.28318 & 0.0001\\
			& -&\textbf{0.317381}&-& 0.408897 & 0.408897 & 3.14159 & -6.8344e-8 \\
			& -&-&\textbf{0.391766}& 0.628319 & 0.628348 & 0.01 & 3.14856 \\
			10 &  \textbf{0.277618}&-&-& 1.8875 & 1.2541 & 6.28318 & 0.0001\\
			& -&\textbf{0.277618}&-& 0.316701 & 0.316701 & 3.14159 & 5.8783e-7  \\
			& -&-&\textbf{0.370164}& 0.628443 & 0.784857 & 0.756098 & 4.60592 \\
			\label{tab:PQStable}
		\end{tabular}
	\end{ruledtabular}
\end{table*}

\begin{figure}
	\centering
	\subfloat[][]{\includegraphics[width=0.99\linewidth,keepaspectratio]{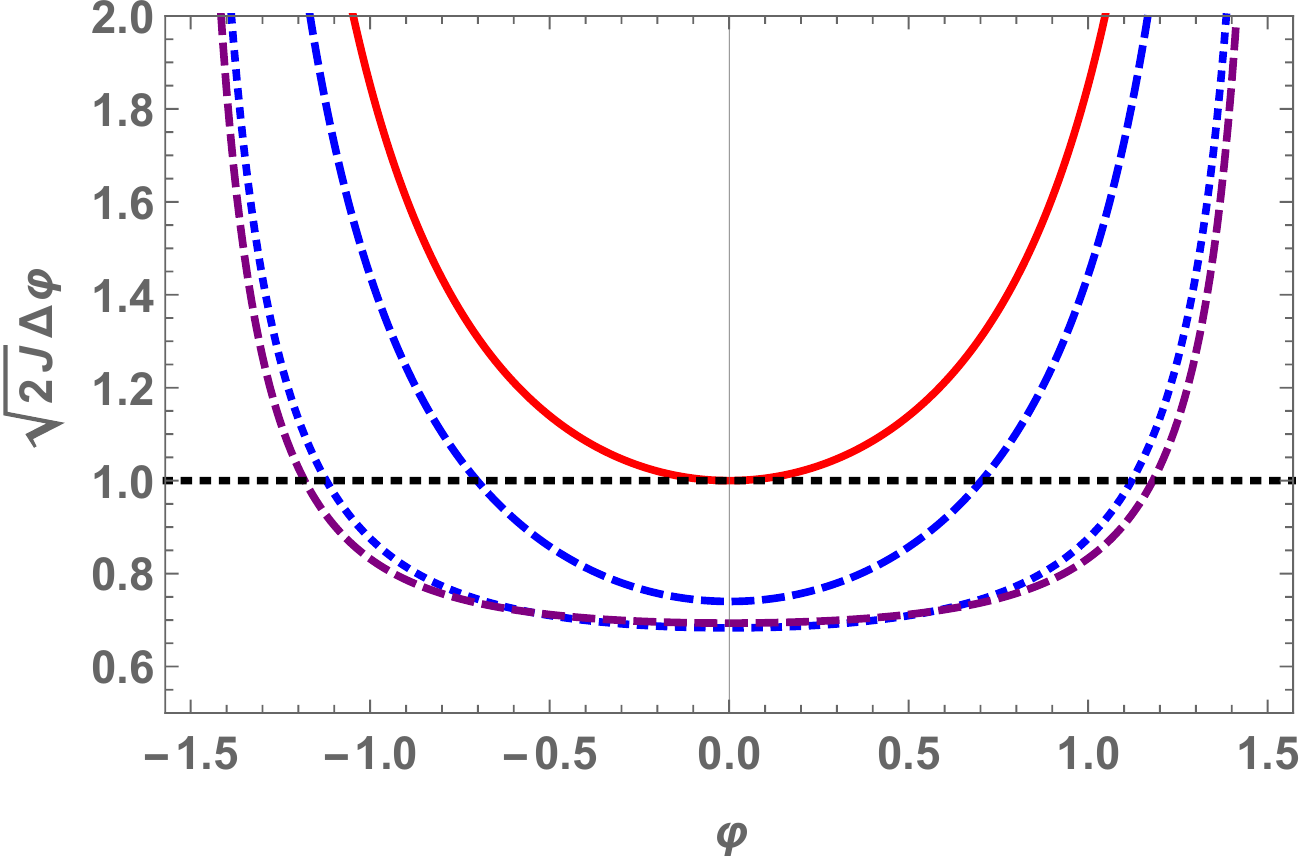}
		\label{fig:PU}}
	\\
	\subfloat[][]{\includegraphics[width=0.99\linewidth,keepaspectratio]{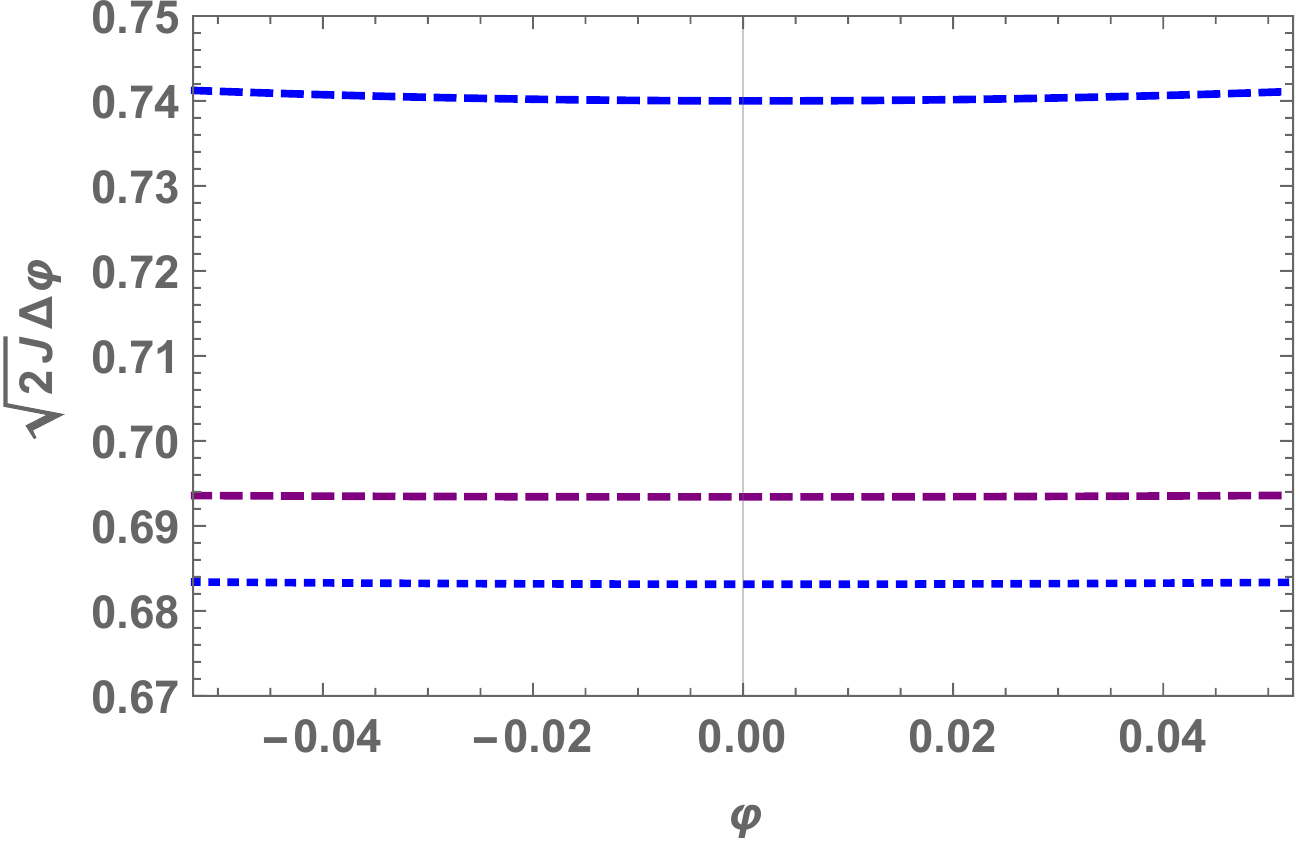}
		\label{fig:PU2}}
	\caption{\ref{fig:PU} Scaled phase uncertainty $\Delta\varphi/\Delta\varphi_{\text{ACS}}$ plotted against phase $\varphi$ with $j=10$. From highest to lowest curve at the origin: atomic coherent state (red, solid), minimum $\xi_{y}^{2}$ spin-squeezed state (blue, dashed), minimum $\xi_{yz}^{2}$ planar-squeezed state (purple, dashed) and minimum $\xi_{W,y}^{2}$ spin-squeezed state (blue, dotted). \ref{fig:PU2} same as Fig.~\ref{fig:PU}, but for $\varphi\sim 0$.  The minimum $\xi_{W,y}^{2}$ spin-squeezed state yields greater sensitivity over the minimum $\xi_{yz}^{2}$ planar-squeezed state for small phase, but is slightly less broad.}
	\label{fig:PU_plots} 
\end{figure}

\section{\label{sec:Metrology} III. A comparison of quantum- enhanced metrological precision between optimally spin- and planar-squeezed States}

\noindent Here we provide a brief idealized comparison of the performance between the maximally spin-squeezed state and the maximally planar-squeezed state in multi-atom spectroscopy where the $\pi/2$-pulses in the Ramsey procedure can be characterized by rotations about the $\pm y$-axis while the free evolution between pulses during which a phase $\varphi$ is acquired can be described by a rotation about the $z$-axis. We take a projection on the $y$-axis as our means of detection.  We note that the optimization performed in this section is with respect to the initial superposition state; not the phase uncertainty itself.  For a given estimator, the optimized (minimum) phase uncertainty is known as the Cram\'{e}r-Rao bound (CRB), or lower-bound on phase estimation, defined in terms of the classical Fisher information $\Delta\phi_{\text{CRB}}= 1/F\left(\phi\right)$, where the Fisher information is given by $F\left(\phi\right)=\sum_{\epsilon}P^{-1}\left(\epsilon|\phi\right)\left(\delta_{\phi}P\left(\epsilon|\phi\right)\right)^{2}$ and where the sum extends over all measurement outcomes $\epsilon$ and where $P\left(\epsilon|\phi\right)$ is the probability of a particular measurement outcome. Optimization of the Fisher information over all possible postive-operator valued measures (POVMs) yields the quantum Fisher information, most often expressed for pure states as $F_{Q}=4\left\{\braket{\delta_\phi \psi|\delta_\phi \psi} - |\braket{\delta_\phi \psi|\psi}|^{2}\right\}$, where $\ket{\psi}$ represents the initial state of the system. The corresponding quantum Cram\'{e}r-Rao upper bound (qCRB) on phase estimation is then $\Delta\phi_{\text{qCRB}} = 1/F_{Q}$, representing the smallest phase uncertainty for a given initial state and transformation, independent of estimation method.  Furthermore, for a given detection observable, optimization of the spin-squeezed parameter as defined by Wineland \textit{et al.} \cite{ref:Wineland1} and given in Eq.~\ref{eqn:Intro_6a}, would yield a phase uncertainty that, by definition, would necessarily coincide with a result obtained through direct optimization of the phase uncertainty. This result, however, does not generally coincide with the QCRB. For a more thorough discussion on the bounds of phase estimation, see Pezz\'{e} \textit{et al.} \cite{ref:Pezze2} and references therein. \\

\noindent The full Ramsey sequence can  be described in terms of SU(2) rotations as

\begin{equation}
	\ket{\text{out}} = e^{-i\tfrac{\pi}{2}\hat{J}_{y}}e^{-i\varphi\hat{J}_{z}}e^{i\tfrac{\pi}{2}\hat{J}_{y}}\ket{\text{in}} = e^{-i\varphi\hat{J}_{x}}\ket{\text{in}},
	\label{eqn:PU1}
\end{equation}

\noindent where we have used $e^{-i\tfrac{\pi}{2}\hat{J}_{y}}\hat{J}_{z}e^{i\tfrac{\pi}{2}\hat{J}_{y}}=\hat{J}_{x}$ \cite{ref:Yurke}. Working in the Heisenberg picture, and using $\hat{J}_{y}^{\text{out}}=e^{i\varphi\hat{J}_{x}}\hat{J}_{y}^{\text{in}}e^{-i\varphi\hat{J}_{x}}$, it can be shown

\begin{align}
	\braket{\hat{J}_{y}^{\text{out}}} &= \braket{\hat{J}_{y}^{\text{in}}}\cos\varphi - \braket{\hat{J}_{z}^{\text{in}}}\sin\varphi, \\
	\nonumber \\
	\braket{\hat{J}_{y}^{2\;\text{out}}} &= \braket{\hat{J}_{y}^{2\;\text{in}}}\cos^{2}\varphi + \braket{\hat{J}_{z}^{2\;\text{in}}}\sin^{2}\varphi- \nonumber \\
	& \;\;\;\;\;\;\;\;\;\;\;\;\;\;\;\;\;\;\;\;\;\;\;\;- \tfrac{1}{2}\sin2\varphi\braket{\big[\hat{J}_{y}^{\text{in}},\hat{J}_{z}^{\text{in}}\big]_{+}},
	\label{eqn:PU2}
\end{align}

\noindent where $\big[\hat{A},\hat{B}\big]_{+}$ denotes the anti-commutator $\hat{A}\hat{B}+\hat{B}\hat{A}$. From these, the variance in $\hat{J}_{y}^{\text{out}}$ can be found in terms of quantities measured with respect to the input state as

\begin{align}
	\Delta^{2}\hat{J}_{y}^{\text{out}} = \Delta^{2}\hat{J}_{y}^{\text{in}}\cos^{2}\varphi + &\Delta^{2}\hat{J}_{z}^{\text{in}}\sin^{2}\varphi \;+\; \nonumber \\
	& +\; \sin2\varphi\;\text{Cov}\big(\hat{J}_{y}^{\text{in}},\hat{J}_{z}^{\text{in}}\big), 
	\label{eqn:PU3}
\end{align}

\noindent where $\text{Cov}\big(\hat{J}_{y}^{\text{in}},\hat{J}_{z}^{\text{in}}\big)$ is the covariance $\text{Cov}\big(\hat{A},\hat{B}\big) = \tfrac{1}{2}\big(\braket{\big[\hat{A},\hat{B}\big]_{+}}-2\braket{\hat{A}}\braket{\hat{B}}\big)$.  The phase uncertainty can be found through the usual error propagation calculus $\Delta\varphi = \Delta\hat{\mathcal{O}}/|\partial_{\varphi}\braket{\hat{\mathcal{O}}}|$, where $\hat{\mathcal{O}}$ denotes a general detection observable,  to yield

\begin{equation}
	\Delta\varphi = \frac{\Delta\hat{J}_{y}^{\text{out}}}{|\partial_{\varphi}\braket{\hat{J}_{y}^{\text{out}}}|} \stackrel{\varphi\;\text{small}}{\approx} \frac{\Delta\hat{J}_{y}^{\text{in}}}{|\braket{\hat{J}_{z}^{\text{in}}}|},
	\label{eqn:PU4}
\end{equation}

\noindent which can be computed numerically for arbitrary phase using Eqs.~\ref{eqn:PU2} and \ref{eqn:PU3}.  As our goal in this exercise is to compare the phase sensitivities obtained using an optimally spin-squeezed state and an optimally planar-squeezed state, we need to be mindful when choosing which spin-direction and plane of squeezing works best for the detection observable and Ramsey procedure we have chosen.  Since planar squeezing has been demonstrated to yield enhanced phase sensitivity due to rotations about the axes perpendicular to the plane of squeezing \cite{ref:He}, we are interested in optimizing $\xi_{yz}^{2}$. For spin-squeezing, we use the form defined by Wineland \textit{et al.} \cite{ref:Wineland1}, and we are interested in optimizing with respect to the $y$-direction: $\xi_{W,y}^{2}$. For $N=2J=20$ atoms, the state optimizing both of these parameters correspond to a cat state with nearly similar angle values (for the SS state, $\theta_1 \sim -0.35$ and $\phi\sim\Phi_r\sim0$) and a MSD almost entirely along the $z$-axis such that $|\braket{\vec{J}_{\parallel}}|\sim \big|\braket{\hat{J}_{z}}\big|\sim Nj$ where $j=1/2$ is the spin value of a single atom in the ensemble. We note here that the emphasis of this idealized example is in comparing the relative phase sensitivity obtained for the optimized spin- and planar-quantum squeezed states without the effects of decoherence and dissipation, which represents a substantial hurdle to the experimental implementation of such states.  The state statistics and properties of the cat states have been studied from a theoretical perspective \cite{ref:Gerry,ref:Puri} as well as in the context of metrology \cite{ref:Lee1} wherein the cat state performance was studied under the effects of dissipation.  The authors of \cite{ref:Lee1} also discuss experiments in which spin cat states have been generated through the use of nonlinear Kerr effects due to atomic collisions in Bose-Josephson systems through dynamical evolution \cite{ref:Ferrini,ref:Pawlowski,ref:Spehner} or ground-state preparation \cite{ref:Lee2,ref:Lee3}. They go on to point out that in the case of no detuning, the self-trapped ground states for Bose-Josephson systems with negative nonlinearity are similar to spin cat states (or more specifically, a macroscopic superposition of spin coherent states of the form $\ket{\Psi}_{M}\propto \ket{\theta,\phi}+\ket{\pi-\theta,\phi}$, where $\ket{\theta,\phi}\Leftrightarrow \ket{\zeta,j}$ with $\zeta=e^{i\phi}\tan{\tfrac{\theta}{2}}$).  We also include the case in which we optimize spin-squeezing along the $y$-direction as per Eq. \ref{eqn:Intro_6} with state parameters taken from Table \ref{tab:SStable}; this corresponds to a state that is neither a cat nor phase cat state. Interestingly, optimization with respect to this particular form of spin-squeeze parameter does not perform as well for small phases as the state obtained through optimization of the Wineland \textit{et al.} \cite{ref:Wineland1} spin-squeeze parameter.  \\

\noindent We plot in Fig.~\ref{fig:PU_plots} the scaled phase uncertainty $\Delta\varphi/\Delta\varphi_{\text{ACS}}=\sqrt{2J}\Delta\varphi$ against phase $\varphi$.  For reference we include a curve corresponding to the phase uncertainty for an atomic coherent state $\ket{\zeta,J=10}$.  Curves falling below $1$ correspond to sub-SQL phase sensitivity. Looking at small phase values in Fig.~\ref{fig:PU2}, the optimally $\xi_{W,y}^{2}$ spin-squeezed states yields greater phase sensitivity over the planar-squeezed state but the planar-squeezed states yield sub-SQL sensitivity over a slightly larger domain of phase values.  The curve corresponding to $\xi_{y}^{2}$ spin-squeezing, while providing enhanced sub-SQL uncertainty for small phases, under-performs compared to the other forms of squeezing both in maximal sensitivity achieved as well as the size of the phase-domain yielding sub-SQL sensitivity. The difference in performance between the optimally spin- and planar-quantum squeezed states  is minor, in this case, as the corresponding states are very similar.  However, this demonstration is meant only to show the characteristic difference between spin-squeezed and PQS states, that is, that SS states are better suited for measuring small changes to phases that are more-or-less known (that is, \textit{a priori} knowledge is assumed; local phase estimation) while PQS states are better suited for measuring phases wherein there is a complete lack of knowledge about the phase (i.e. a ``global" phase).  

\section{\label{sec:SpinSqueeze_Conclusion} IV. Conclusion}

\noindent In this paper we discussed the presence of spin- and planar- squeezing in generalized superpositions of atomic coherent states by optimizing over the state parameters that determine their individual locations on the Bloch sphere as well as the relative phase between the constituent components in the superposition state.  We included a brief review describing the use of these forms of squeezing as a means of gauging sub-SQL phase sensitivity in Ramsey spectroscopy as well as a means of constructing an entanglement witness and determining the depth of entanglement.  For the former point, the presence of spin-squeezing necessitates entanglement within the system and leads to sub-SQL sensitivities.  The same is not true of planar-quantum squeezed states, in which the system may not be entangled (for example, the ACS); entanglement is still required to yield sub-SQL sensitivities, which can be confirmed through calculation of the ``depth of entanglement", which requires computation of the planar-squeeze parameter once for a given value of $J$.  We have shown that the states that maximize spin- (planar-) squeezing along a particular direction (plane) are often closely related to the so-called phase cat states in which the states comprising the superposed state have the same displacement from the $z$-axis but are oriented differently in the $x-y$ plane (i.e. $\theta_{1}=\theta_{2},\;\phi_{1}\neq\phi_{2}$).   We included a brief review on planar squeezing as well as the different forms of spin-squeezing discussed in the literature, most notably as it is defined by Wineland \textit{et al.} \cite{ref:Wineland1} as it pertains to metrological precision and S{\o}renson \textit{et al.} \cite{ref:Sorenson2} as it pertains to a multi-partite separability criterion.  \\

\noindent Finally, we closed with an application to Ramsey spectroscopy wherein we parameterize the superposition state of Eqn.~\ref{eqn:Intro_4} in different ways to optimize the presence of two different forms of spin-squeezing (as defined by Wineland \textit{et al.} \cite{ref:Wineland1} and S{\o}renson \textit{et al.} \cite{ref:Sorenson2}) as well as planar-quantum squeezing. Due to the proportional relationship between the phase uncertainty and the Wineland \textit{et al.} \cite{ref:Wineland1} spin-squeeze parameter, optimization of spin-squeezing is equivalent to directly optimizing the phase uncertainty. This is not generally true for the S{\o}renson \textit{et al.} \cite{ref:Sorenson2} definition of spin-squeezing nor for planar-quantum squeezing. We have shown, in an idealized comparison, that the SS superposition state yields slightly greater sensitivity over the PQS state for small phases, however the PQS yields sub-SQL sensitivty over a slightly larger phase-domain.  While not a drastic demonstration, this supports the commonly held notation that SS states are better suited for local parameter estimation while PQS are better suited for parameter estimation when \textit{a priori} knowledge exists.
    
\section{\label{sec:SpinSqueeze_Acknowledgements} V. Acknowledgments}

\noindent RJB acknowledges support from the National Research Council Research Associate Program (NRC RAP). EEHIII and CCG acknowledge support under AFRL Summer Faculty Fellowship  Program (SFFP). EEHIII and JZ acknowledges support under the AFRL Material Command Grant FA8750-16-2-0140.
PMA and CCG acknowledge support from the Air Force Office of Scientific Research (AFOSR).  
Any opinions, findings and conclusions  or  recommendations  expressed  in  this  material are those of the author(s)
and do not necessarily reflect the views of the Air Force Research Laboratory (AFRL).

\nocite{apsrev41Control}
\bibliographystyle{apsrev4-1}
\bibliography{SpinSqueeze}% Produces the bibliography via BibTeX.

\end{document}